\documentstyle[aps,prb,twocolumn,epsf,floats]{revtex}
\begin{document}
\draft
\preprint{}
\twocolumn[\hsize\textwidth\columnwidth\hsize\csname
@twocolumnfalse\endcsname
\title{
     Resonant inelastic x-ray scattering in one-dimensional copper oxides
}
\author{
                K. Tsutsui, T. Tohyama, and S. Maekawa
}
\address{
                       Institute for Materials Research,
                  Tohoku University, Sendai 980-8577, Japan
}
\date{Received 23 September 1999}
\maketitle
\begin{abstract}
The Cu $K$-edge resonant inelastic x-ray scattering (RIXS) spectrum in
 one-dimensional insulating cuprates is theoretically examined by using
 the exact diagonalization technique for the extended one-dimensional
 Hubbard model with nearest-neighbor Coulomb interaction.
We find the following characteristic features that can be detectable
 by RIXS experiments:
(i) The spectrum with large momentum transfer indicates the formation
 of excitons, i.e. bound states of {\it holon} and {\it doublon}.
(ii) The spectrum with small momentum transfer depends on the incident
 photon energy.
We propose that the RIXS provides a unique opportunity to study the upper
 Hubbard band in one-dimensional cuprates.
\end{abstract}

]
\narrowtext

Insulating copper oxides, when mapped onto a single-band component,\cite{Zhang}
 have their occupied lower Hubbard band (LHB) and the unoccupied upper
 Hubbard band (UHB) separated by a Mott gap.
Angle-resolved photoemission spectroscopy (ARPES) has succeeded in observing
 the momentum dependence of the LHB for two-dimensional (2D) (Ref.~2)
 %%%%%%%%%%%%~\cite{Wells}
 and 1D (Ref.~3)
 %%%%%%%%%%%%~\cite{Kim}
 cuprates.
In contrast to the LHB, the UHB is far from our understanding.

The information about the UHB can be extracted from excitations from the LHB to
UHB across the Mott gap.
A representative technique for this is electron energy-loss spectroscopy
 (EELS), which has been applied to both 2D (Ref.~4)
 %%%%%%%%%%%\cite{Wang}
 and 1D (Ref.~5)
 %%%%%%%%%%%\cite{Neudert}
 insulating cuprates.
Another technique is resonant inelastic x-ray scattering (RIXS).
When the incident photon energy $\omega_i$ is tuned through the Cu $K$
 absorption edge, the momentum dependence of the excitations from the LHB to
 UHB has been observed for insulating 2D cuprates, Sr$_2$CuO$_2$Cl$_2$
 (Ref.~6)
 %%%%%%%%%%%~\cite{Abbamonte}
 and Ca$_2$CuO$_2$Cl$_2$ (Ref.~7).
 %%%%%%%%%%%~\cite{Hasan}
On the theoretical side, the present authors predicted the
 dependence based on a 2D Hubbard model with second- and third-neighbor
 hopping terms.\cite{Tsutsui}

Typical examples of 1D cuprates are Sr$_2$CuO$_3$ and SrCuO$_2$.
Their magnetic susceptibilities are explained by the 1D Heisenberg
 model.\cite{Motoyama}
ARPES measurements~\cite{Kim,Fujisawa} showed that a photodoped
 hole cannot exist as a quasiparticle, but changes into two collective
 excitations, a {\it spinon} and a {\it holon}, as expected from the 1D
 single-band Hubbard and $t$-$J$ models.
From EELS experiments,\cite{Neudert} it has been pointed out that an
 extended version of the Hubbard model with a moderate nearest-neighbor
 Coulomb interaction $V$ is necessary to understand excitations across
 the Mott gap.
It is interesting to know what we obtain in the RIXS spectrum prior to
 the experiments.

In this study, the momentum dependence of the RIXS spectrum in 1D cuprates
 is demonstrated by using the exact diagonalization technique for the
 extended Hubbard model.
We find the following characteristic features that can be detectable by
 RIXS experiments:
The spectral weight concentrates on the narrow energy region due to the
 formation of excitons.
This is also seen in the dynamical charge-correlation function, although the
 dependence of the intensity on the momentum transfer is different.
We also find that the spectral weight distribution for a small momentum
 transfer depends on the incident photon energy.
This is associated with the intermediate state that is characteristic
 of the RIXS process.

A minimal model that can describe the LHB and UHB is the extended
 Hubbard model as used in the analyses of the EELS spectra.~\cite{Neudert}
Hereafter, we use the term ``3$d$-electron system'' to stand for the LHB and UHB.
The Hamiltonian is given by
\begin{eqnarray}
\label{ham3d}
H_{3d} =&& -t\sum_{i, \sigma} (d_{i,\sigma}^\dag d_{i+1,\sigma} + {\rm H.c.})
      +U\sum_i n^d_{i,\uparrow}n^d_{i,\downarrow} \nonumber\\
      &&+V\sum_i n^d_i n^d_{i+1},
\end{eqnarray}
where $d_{i,\sigma}^\dag$ is the creation operator of $3d$ electron with
 spin $\sigma$ at site $i$, $n_{i,\sigma}^d$=$d_{i,\sigma}^\dag d_{i,\sigma}$,
and $n^d_i$=$n^d_{i,\uparrow}+n^d_{i,\downarrow}$.
The on-site Coulomb energy $U$ corresponds to the charge-transfer energy
 of cuprates.\cite{Tsutsui}

In the intermediate states of the Cu $K$-edge RIXS process, 3$d$ electrons
 interact with a 1$s$-core hole created by the dipole transition of an
 1$s$ electron to 4$p$ orbital due to absorption of an incident photon
 with energy $\omega_i$ and momentum ${\bf K}_i$.
This interaction is written as
\begin{eqnarray}
\label{ham1s3d}
H_{1s\text{-}3d}=-V_c\sum_{i,\sigma,\sigma'} n_{i,\sigma}^d n_{i,\sigma'}^s,
\end{eqnarray}
where $n_{i,\sigma}^s$ is the number operator of the 1$s$-core hole with spin
 $\sigma$ at site $i$, and $V_c$ is taken to be positive.
This interaction $V_c$ causes excitations of 3$d$ electrons across
 the gap.
The photoexcited 4$p$ electron is assumed to go into the bottom of the
 4$p$ band with momentum ${\bf k}_0$ and not to interact with either the
 3$d$ electrons or 1$s$-core hole due to the delocalized nature of the
 4$p$ orbital.\cite{Tsutsui}
In the final state, the 4$p$ electron goes back to the 1$s$ orbital,
 emitting a photon with energy $\omega_f$ and momentum ${\bf K}_f$.
The RIXS spectrum is then given by~\cite{Tsutsui}
\begin{eqnarray}
\label{rixs}
I(\Delta {\bf K}&&,\Delta\omega)=\sum_\alpha\left|\langle\alpha|
\sum_\sigma s_{{\bf k}_0-{\bf K}_f,\sigma} p_{{\bf k}_0,\sigma}
\right.\nonumber\\\times&&\left.
\frac{1}{H+\varepsilon_{1s\text{-}4p}-E_0-\omega_i-i\Gamma}
p_{{\bf k}_0,\sigma}^\dag s_{{\bf k}_0-{\bf K}_i,\sigma}^\dag
|0\rangle\right|^2
\nonumber\\\times&&
\delta(\Delta\omega-E_\alpha+E_0),
\end{eqnarray}
where $H$=$H_{3d}$+$H_{1s\text{-}3d}$, $\Delta{\bf K}$=${\bf K}_i-{\bf K}_f$,
 $\Delta\omega$=$\omega_i-\omega_f$, $s_{{\bf k},\sigma}^\dag$
 ($p_{{\bf k},\sigma}^\dag$) is the creation operator of the 1$s$-core hole
 (4$p$ electron) with momentum ${\bf k}$ and spin $\sigma$, $|0\rangle$ is
 the ground state of the half-filled system with energy $E_0$,
 $|\alpha\rangle$ is the final state of the RIXS process with energy
 $E_\alpha$, $\Gamma$ is the inverse of the relaxation time in the
 intermediate state, and $\varepsilon_{1s\text{-}4p}$ is the energy
 difference between the $1s$ level and the bottom of the $4p_z$ band.
The momentum component parallel to the 1D Cu chain is represented by
 $\Delta K$ hereafter.
The values of the parameters are set to be $U/t$=10, $V_c/t$=15, and
 $\Gamma/t$=1 as for the 2D cuprates.\cite{Tsutsui}
The RIXS spectrum in Eq.~(\ref{rixs}) is calculated on a 14-site ring by
 using a modified version of the conjugate-gradient method together with
 the Lancz\"os technique.

%%%%%%%%%%%%%%%%%%%%%%%%%%%%%%%%%%%%%%%%%%%%%%%%%%%%%%%%%%%%%%%%%%%%%%%%%%%%%%%
\begin{figure}
%\vskip7cm
\epsfxsize=7cm
\centerline{\epsffile{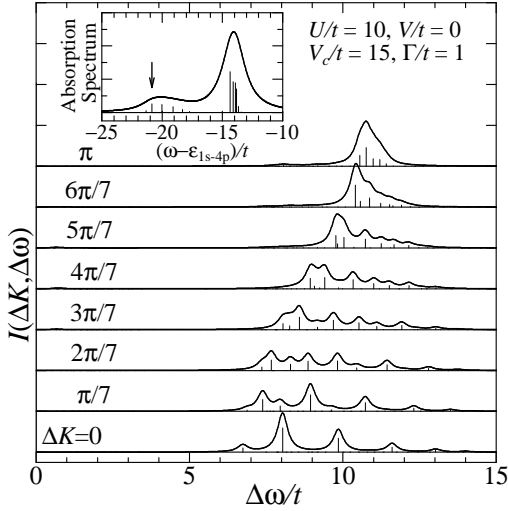}}
\caption{
Resonant inelastic x-ray scattering spectra for Cu $K$-edge in the 1D
 half-filled Hubbard model.
The spectra of the elastic scattering process at $\Delta K$=0 are not shown.
The parameters used are $U/t$=10, $V/t$=0, $V_c/t$=15, and $\Gamma/t$=1.
The $\delta$ functions (the vertical thin solid lines) are convoluted
 with a Lorentzian broadening of 0.2$t$.
The inset is the Cu 1$s$ absorption spectra with a broadening of 1.0$t$, and the
 incident photon energy $\omega_i$ is set to the value denoted by the arrow.
}
\label{figrixsV0}\end{figure}
%%%%%%%%%%%%%%%%%%%%%%%%%%%%%%%%%%%%%%%%%%%%%%%%%%%%%%%%%%%%%%%%%%%%%%%%%%%%%%%

\begin{figure}
%\vskip7cm
\epsfxsize=7cm
\centerline{\epsffile{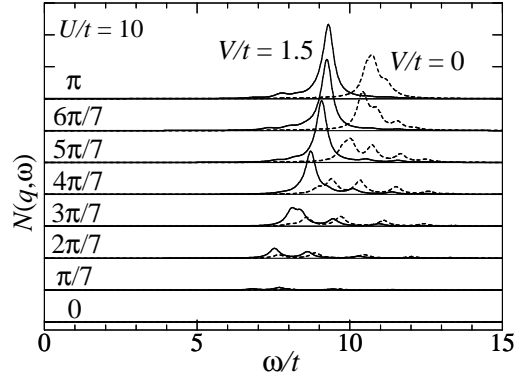}}
\caption{
Dynamical charge correlation function $N(q,\omega)$ in the 1D half-filled
 Hubbard model.
The parameters used are $U/t$=10, $V/t$=0 (broken lines), and $V/t$=1.5
 (solid lines).
The $\delta$-functions are convoluted with Lorentzian broadening of 0.2$t$.
}
\label{fignqw}\end{figure}

Figure~\ref{figrixsV0} shows the momentum dependence of the RIXS spectrum
 for the case of $V/t$=0.
The inset shows the absorption spectrum of the 1$s$ electron to the 4$p$
 orbital.\cite{Tsutsui}
The incident photon energy $\omega_i$ that appears in Eq.~(\ref{rixs}) is
 set to be the value denoted by the arrow.
This means that the states near the arrow satisfy the resonance conditions
 of Eq.~(\ref{rixs}).
For small $\Delta K$, the spectra spread over about 8$t$.
The energy region, however, shrinks with increasing $\Delta K$, and the
 spectral weight concentrates on a narrow energy region at $\Delta K$=$\pi$.
This momentum dependence is similar to that of the dynamical
 charge-correlation function $N(q,\omega)$, shown by dashed lines
 in Fig.~\ref{fignqw}, which is explained by the particle-hole model
 with only a charge degree of freedom.\cite{Neudert,Stephan}
In the particle-hole model, {\it holon} and {\it doublon} bands separated
 by $U$ describe the LHB and UHB, respectively.
We note that the spin degree of freedom is less effective on $N(q,\omega)$
 because of the wave-function factorization~\cite{Ogata} in the large-$U$
 limit.
At the zone center of the interband transitions, one obtains a continuum with
 a width of 8$t$ above the gap of $U$$-$$4t$.
The width of the continuum decreases with an increase of the momentum transfer,
 and finally becomes an excitation with energy $U$ at the zone boundary.
This momentum dependence is qualitatively consistent with that shown in
 Fig.~\ref{figrixsV0} and also with $N(q,\omega)$ shown in Fig.~\ref{fignqw}.
It is interesting to note that the integrated weight of
 $I(\Delta K,\Delta \omega)$, with respect to $\Delta\omega$, is almost
 independent of $\Delta K$ in contrast to $N(q,\omega)$ where the
 integrated  weight is proportional to $\sin^2(q/2)$
 for the large-$U$ region.\cite{Stephan}

%%%%%%%%%%%%%%%%%%%%%%%%%%%%%%%%%%%%%%%%%%%%%%%%%%%%%%%%%%%%%%%%%%%%%%%%%%%%%%%
%%%%%%%%%%%%%%%%%%%%%%%%%%%%%%%%%%%%%%%%%%%%%%%%%%%%%%%%%%%%%%%%%%%%%%%%%%%%%%%

\begin{figure}
%\vskip7cm
\epsfxsize=7cm
\centerline{\epsffile{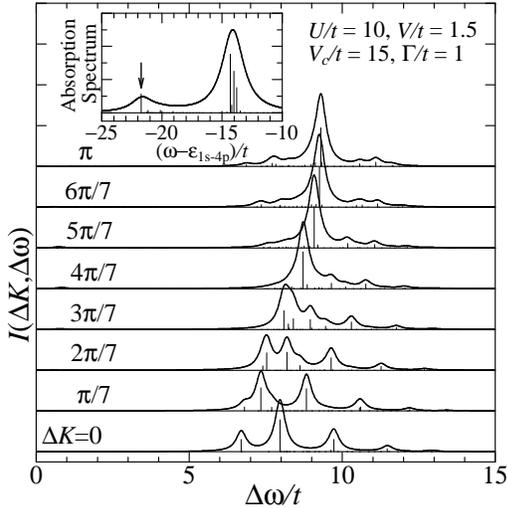}}
\caption{
The same as Fig.~\ref{figrixsV0} but for $V/t$=1.5.}
\label{figrixsV15}\end{figure}

The effect of the nearest-neighbor Coulomb interaction $V$ on
 $I(\Delta K,\Delta \omega)$ is shown in Fig.~\ref{figrixsV15}, where
 $V/t$=1.5 is taken according to the analyses of the EELS data.\cite{Neudert}
From the comparison between Figs.~\ref{figrixsV0} and \ref{figrixsV15},
 we find that the shape of the spectrum changes remarkably for
 $\Delta K$$>$$\pi/2$, accompanied by the formation of sharp peaks
 together with the shift of the spectral weight to the lower-energy region.
This is due to excitons that appear at the momenta satisfying a condition
 that $V$$>$$2t\cos (\Delta K/2)$.~\cite{Stephan}
A similar change is also seen in $N(q,\omega)$ (see Fig.\ref{fignqw}).
For $\Delta K$$<$$\pi/2$, the intensity of the spectrum near the
 lower edge is enhanced by the presence of excitonic effects, which is
 again similar to the feature of $N(q,\omega)$.\cite{Stephan}
We note that the presence of $V$ also affects the RIXS spectrum in the
 2D case, although the edge structure, which is a fingerprint of the
 shape of the UHB,~\cite{Tsutsui} does not depend on $V$.

\begin{figure}
%\vskip7cm
\epsfxsize=7cm
\centerline{\epsffile{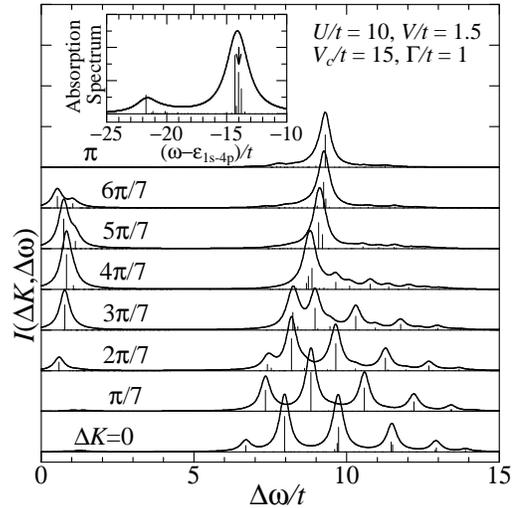}}
\caption{
The same as Fig.~\ref{figrixsV15}, but with the incident photon
 energy $\omega_i$ set to the value denoted by the arrow in the inset.
}
\label{figrixsV15e14}\end{figure}

The spectral weight depends on the incident photon energy $\omega_i$.
By changing $\omega_i$$-$$\varepsilon_{1s\text{-}4p}$ from about $-21t$
 to $-14t$, the distribution of intensity in the lower-momentum region
 shifts to the higher-energy region, as shown in Fig.~\ref{figrixsV15e14}.
This can be attributed to the nature of the intermediate state around
 the resonance condition, denoted by the arrows in the insets of
 Figs.~\ref{figrixsV0} and \ref{figrixsV15}.
The peaks at $\omega$$-$$\varepsilon_{1s\text{-}4p}$$\sim$$-21t$
 ($\sim U$$-$2$V_c$) in the absorption spectrum are characterized by
 states having configurations with the core-hole site doubly occupied
 by 3$d$ electrons together with an empty site of 3$d$ electrons.
On the other hand, in states near
 $\omega$$-$$\varepsilon_{1s\text{-}4p}$=$-14t$ ($\sim -V_c$), the core-hole
 site is singly occupied by 3$d$ electrons and there is no empty site.
Since the different intermediate states are used depending on the incident
 photon energy, the RIXS spectrum depends on the energy.
The spectrum in the low-energy region of $\Delta\omega/t$$\alt$1 in
 Fig.~\ref{figrixsV15e14} corresponds to singlet excitations of the spinon
 pair, because the intermediate state has singly occupied configurations.

%%%%%%%%%%%%%%%%%%%%%%%%%%%%%%%%%%%%%%%%%%%%%%%%%%%%%%%%%%%%%%%%%%%%%%%%%%%%%%%
%%%%%%%%%%%%%%%%%%%%%%%%%%%%%%%%%%%%%%%%%%%%%%%%%%%%%%%%%%%%%%%%%%%%%%%%%%%%%%%

In summary, we have examined the momentum dependence of the RIXS spectrum
 in 1D cuprates, which provides a unique opportunity to study the upper
 Hubbard band, using the exact diagonalization technique
 for the extended Hubbard model.
We have found the following characteristic features of the RIXS spectrum:
The spectrum with large momentum transfer, $\Delta K$$>$$\pi/2$,
 indicates the formation of excitons, i.e., bound states of
 a {\it holon} and a {\it doublon}.
We have also found that the spectral weight distribution for small
 momentum transfer, $\Delta K$$<$$\pi/2$, depends on the incident
 photon energy.
The dependence is associated with the intermediate state that is
 characteristic of the RIXS process.
These findings will be good guides for RIXS experiments on 1D
 insulating cuprates that will be performed in the future.

%%%%%%%%%%%%%%%%%%%%%%%%%%%%%%%%%%%%%%%%%%%%%%%%%%%%%%%%%%%%%%%%%%%%%%%%%%%%%%%
%%%%%%%%%%%%%%%%%%%%%%%%%%%%%%%%%%%%%%%%%%%%%%%%%%%%%%%%%%%%%%%%%%%%%%%%%%%%%%%

This work was supported by Priority-Area Grants from the Ministry of
 Education, Science, Culture and Sport of Japan, CREST, and NEDO.
Computations were carried out in ISSP, University of Tokyo; IMR,Tohoku
 University; and Tohoku University.

\end{document}